\documentclass[twocolumn,superscriptaddress,nofootinbib,aps,pre,10pt]{revtex4-1}

\usepackage{amsmath,amssymb,graphicx,epstopdf,mathtools,color,url,siunitx}
\usepackage{physics}

\makeatletter 
\renewcommand{\pdv}[2]{\begingroup 
  \@tempswafalse\toks@={}\count@=\z@ 
  \@for\next:=#2\do 
    {\expandafter\check@var\next\@nil
     \advance\count@\der@exp 
     \if@tempswa 
       \toks@=\expandafter{\the\toks@}%
     \else 
       \@tempswatrue 
     \fi 
     \toks@=\expandafter{\the\expandafter\toks@\expandafter\partial\der@var}}%
  \frac{\partial\ifnum\count@=\@ne\else^{\number\count@}\fi#1}{\the\toks@}%
  \endgroup} 
\def\check@var{\@ifstar{\mult@var}{\one@var}} 
\def\mult@var#1#2\@nil{\def\der@var{#2^{#1}}\def\der@exp{#1}} 
\def\one@var#1\@nil{\def\der@var{#1}\chardef\der@exp\@ne} 
\makeatother 

\allowdisplaybreaks

\begin{document}
\title{How to measure heat in stochastic systems}
\author{D. Chiuchi\`u}
\email{davide.chiuchiu@nipslab.org}
\affiliation{NiPS Lab, Universit\`a degli studi di Perugia, Dipartimento di Fisica e Geologia}

\begin{abstract}
Heat is a complex quantity to measure in stochastic systems because it requires extremely small sampling timesteps. Unfortunately this is not always possible in real experiments, mainly because experimental setups have technical limits. To overcome this difficulty a Simpson-like quadrature scheme was suggested in [\emph{Phil. Trans. R. Soc. A 2017 375}] as a tool to compute the heat in stochastic systems. In this paper we study this new quadrature scheme. In particular, we first give a qualitative proof of the Simpson-like quadrature with the help of Riemann-Stieltjes integrals and we then perform supplementary numerical simulations to confirm our observations. Our main finding is that the Simpson-like quadrature yields errors that are much smaller than the ones obtained with the Stratonovich quadrature. This opens the possibility to design extremely sensitive experiments on stochastic systems without state-of-the-art sampling techniques. 
\end{abstract}

\maketitle

\section{Introduction}
Stochastic thermodynamics is a branch of nonequilibrium statistical physics that deals with the thermodynamic properties of small physical systems in fluctuating environments \cite{seifert}. Its main applications are in nanotechnology and in biophysics, where it's respectively used to study the physical limits of a computing device \cite{sagawa,jarz,barato2,lopez,horowitz,chiuchiu,madamichiuchiu},  and to study  chemical reaction networks and proteins functionality \cite{polettini,pietzonka,zimmermann,hartich}. One of the most important quantity in stochastic thermodynamics is the heat exchanged by the systems with a thermal reservoir. If we consider a system with a single fluctuating degree of freedom $x$  and total energy $H(x)$, the heat $\Delta_t^{\mathrm{d}t}Q$ exchanged with the reservoir between time $t$ and $t+\mathrm{d}t$ is defined as the stochastic integral \cite{sekimoto,seifert,jarzynski2}
\begin{equation}\label{eq:Stratonovich}
\Delta_t^{\mathrm{d}t}Q=\int_{t}^{t+\mathrm{d}t} F_\xi\, \mathrm{d}x_\xi
\end{equation}
where $\xi$ is a dummy time-variable,  $x_\xi=x(\xi)$ and
\begin{equation}
F_\xi=-\left.\pdv{H}{x}\right|_{x=x_\xi},
\end{equation}
To write eq.\eqref{eq:Stratonovich} we use the Stratonovich rules for stochastic integrals \cite{stratonovich, kuo, gardiner, klodenplaten}, so we can always evaluate it with the middle-point (Stratonovich) quadrature
\begin{equation}\label{eq:Stratonivic_quadrature}
\Delta_{t}^{\mathrm{d}t}Q\approx (x_{t+\mathrm{d}t}-x_t)\ F_{t+\frac{\mathrm{d}t}{2}}.
\end{equation}
As any quadrature scheme, eq.\eqref{eq:Stratonivic_quadrature} fails when $\mathrm{d}t$ becomes sufficiently large. This is not dramatic when we study the mathematical theory of stochastic processes because the formal manipulation of stochastic integrals requires the limit $\mathrm{d}t\to 0$ \cite{stratonovich, kuo, gardiner, klodenplaten}. However, this is a serious problem in experiments: due to the technology limitations that exists in any experimental setup, there is always a minimum allowed size for the sampling interval $\mathrm{d}t$. Unfortunately, the sampling intervals used today in the study of stochastic thermodynamics are such that approximation errors in eq.\eqref{eq:Stratonivic_quadrature} are critical. For example, in \cite{bechhoefer2}, the numerical errors of eq.\eqref{eq:Stratonivic_quadrature} are such that it's not possible to measure the time evolution of the heat exchanged by the system with a reservoir. We are not aware if this same problem affects other experimental setups like for example the ones in \cite{ciliberto,ciliberto2,roldan,gieseler,mestres}; however the similarities between the setups in \cite{ciliberto,ciliberto2,roldan,gieseler,mestres} and the one in \citep{bechhoefer2} suggest us that this is a reasonable working hypothesis \footnote{Note that heat in \cite{ciliberto,ciliberto2} is measured through the first principle of thermodynamics and not with eq.\eqref{eq:Stratonivic_quadrature}. Only this latter measurement can show that numerical divergences do not affect the experimental setup.}. This is a major inconvenient because we have recently shown in \cite{chiuchiu} that the time evolution of thermodynamic observables describes interesting features of the system under study. To overcome this limitation we need a numerical quadrature of eq.\eqref{eq:Stratonovich} with an approximation error smaller than the one from eq.\eqref{eq:Stratonivic_quadrature}. {  Unfortunately, the existing literature on the numerical evaluation of stochastic integrals is scarce at best \cite{badr}, so we do not have a mathematical framework to derive more precise quadrature schemes of eq.\eqref{eq:Stratonovich}. As a consequence, we can only make educated guess based on known algorithms for deterministic functions \cite{numericalrecipes} and see how they perform with numerical examples. One among all the possible educated guess for an  efficient quadrature of eq.\eqref{eq:Stratonovich}} is the Simpson-like quadrature
\begin{equation}\label{eq:Simpson_quadrature}
\Delta_{t}^{\mathrm{d}t}Q\approx \frac{x_{t+\mathrm{d}t}-x_{t}}{6}\ \left[F_{t}+F_{t+\mathrm{d}t}+4\ \widetilde{F}_{t+\frac{\mathrm{d}t}{2}}\right].
\end{equation}
where
\begin{equation}
\widetilde{F}_{t+\frac{\mathrm{d}t}{2}}=-\left.\pdv{H}{x}\right|_{x=\frac{x_t+x_{t+\mathrm{d}t}}{2}}
\end{equation}

Eq.\eqref{eq:Simpson_quadrature} was used for the first time\footnote{Usage of eq.\eqref{eq:Simpson_quadrature} is currently not reported in \cite{chiuchiu}. We will correct this issue with an erratum soon.} in \cite{chiuchiu} and was suggested to the authors of \cite{bechhoefer2} in a private communication. References \cite{bechhoefer2,chiuchiu} shows that eq.\eqref{eq:Simpson_quadrature} works reasonably well in both simulations and experiments, thus overcoming the main problems of eq.\eqref{eq:Stratonivic_quadrature}. {  However, we believe that the full implications of eq.\eqref{eq:Simpson_quadrature} to stochastic methods and stochastic thermodynamics have not been properly discussed yet. 

In this paper we want to address this point and, in particular, we want to show that eq.\eqref{eq:Simpson_quadrature} can significantly boost the number of experiments currently available in the area of Stochastic thermodynamics. To this end we proceed as follows: in Section  \ref{sec:riemann_ste} we first give a qualitative derivation of eq.\eqref{eq:Simpson_quadrature} with Riemann-Stieltjes integrals and deterministic functions. Then we discuss two critical passages that do not trivially extend to stochastic functions in Section \ref{sec:stochastic_gen}. This study involves the numerical simulation of a realistic experimental setup currently used in stochastic thermodynamics, and it conveys the main results of this paper. Section \ref{sec:conclusions} is finally devoted to some conclusions and comments.}

\section{Derivation of the Simpson-like quadrature}\label{sec:riemann_ste}
For the moment we assume that $x_\xi$ and $F_\xi$ are deterministic and that $x_\xi$ is a $\mathcal{C}^2$ function. Within these assumptions our problem is equivalent to say that we want to approximate the Riemann-Stieltjes integral
\begin{equation}\label{eq:Riemann}
\int_t^{t+\mathrm{d}t} F_\xi\ \mathrm{d}x_\xi
\end{equation}
with a three-points formula. 
Following \cite{mercer}, the three-points quadrature  of eq.\eqref{eq:Riemann} is
\begin{equation}\label{eq:Riemann_three_points}
\int_t^{t+\mathrm{d}t} F_\xi\ \mathrm{d}x_\xi\approx A F_t+B F_{t+\mathrm{d}t}+C F_{t+\frac{\mathrm{d}t}{2}}.
\end{equation}
where the coefficients $A$, $B$ and $C$ are chosen in such a way that eq.\eqref{eq:Riemann_three_points} is exact for $F_\xi=1$, $F_\xi=\xi$ and $F_\xi=\xi^2$, which implies
\begin{equation}\label{eq:coefficients}
\begin{aligned}
A=&\left(1+\frac{3t}{\mathrm{d}t}+\frac{2t^2}{\mathrm{d}t^2}\right)(x_{t+\mathrm{d}t}-x_t)-\frac{(4t+3\mathrm{d}t)I_1-2I_2}{\mathrm{d}t^2},\\
B=&\left(\frac{t}{\mathrm{d}t}+\frac{2t^2}{\mathrm{d}t^2}\right)(x_{t+\mathrm{d}t}-x_t)-\frac{(4t+\mathrm{d}t)I_1-2I_2}{\mathrm{d}t^2},\\
C=&-4\left[\left(\frac{t^2}{\mathrm{d}t^2}+\frac{t}{\mathrm{d}t}\right)(x_{t+\mathrm{d}t}-x_t)-\frac{(2t+\mathrm{d}t)I_1- I_2}{\mathrm{d}t^2} \right].
\end{aligned}
\end{equation}
where
\begin{subequations}\label{eq:I12defs}
\begin{align}
I_1=&\int_t^{t+\mathrm{d}t} \xi\ \mathrm{d}x_\xi\\
I_2=&\int_t^{t+\mathrm{d}t} \xi^2 \ \mathrm{d}x_\xi
\end{align}
\end{subequations}
Since  $x_\xi$ is a $\mathcal{C}^2$ function we can (i) perform the substitution $\mathrm{d}x_\xi=\pdv{x}{\xi}\, \mathrm{d}\xi$ in eq.\eqref{eq:coefficients}, (ii) integrate by parts, and (iii) use the Simpson quadrature for scalar functions \cite{numericalrecipes} to compute the remaining integrals. These steps imply
\begin{subequations}\label{eq:inquired_equalities}
\begin{align}
\label{eq:inquired_1}&I_1\approx (t+\mathrm{d}t)\, x_{t+\mathrm{d}t} -t\, x_t - \frac{\mathrm{d}t}{6}\left[x_t+x_{t+\mathrm{d}t}+4x_{t+\frac{\mathrm{d}t}{2}} \right],\\
\label{eq:inquired_2}
&\begin{aligned}
I_2 \approx& (t+\mathrm{d}t)^2\,x_{t+\mathrm{d}t} -t^2\,x_t \\
&-\frac{\mathrm{d}t}{3}\left[t\,x_t +(t+\mathrm{d}t) x_{t+\mathrm{d}t} +2(2t+\mathrm{d}t)x_{t+\frac{\mathrm{d}t}{2}} \right] .
\end{aligned}
\end{align}
\end{subequations}
Upon substituting eqs.\eqref{eq:coefficients}-\eqref{eq:inquired_equalities} in eq.\eqref{eq:Riemann_three_points}, we obtain 
\begin{equation}\label{eq:Simpson_Stieltijes}
\begin{aligned}
\int_t^{t+\mathrm{d}t} F_\xi\ \mathrm{d}x_\xi\approx& \frac{x_{t+\mathrm{d}t}-x_t}{6}\left[F_t+F_{t+\mathrm{d}t}+4F_{t+\frac{\mathrm{d}t}{2}}\right]\\
&+\frac{2}{3} (F_{t+\mathrm{d}t}-F_t) \left[\frac{x_t+x_{t+\mathrm{d}t}}{2}-x_{t+\frac{\mathrm{d}t}{2}} \right]
\end{aligned}
\end{equation}
which is the Simpson rule for Riemann-Stieltjes integrals. It is possible to recast it as the simpler formula 
\begin{equation}\label{eq:Simpson_like}
\int_t^{t+\mathrm{d}t} F_\xi\ \mathrm{d}x_\xi\approx \frac{x_{t+\mathrm{d}t}-x_t}{6}\left[F_t+F_{t+\mathrm{d}t}+4\widetilde{F}_{t+\frac{\mathrm{d}t}{2}}\right]
\end{equation}
upon performing the following approximations
\begin{subequations}\label{eq:inquired_approximations}
\begin{align}
F_{t+\frac{\mathrm{d}t}{2}}&\approx \widetilde{F}_{t+\frac{\mathrm{d}t}{2}}\\
x_{t+\frac{\mathrm{d}t}{2}}&\approx \frac{x_t+x_{t+\mathrm{d}t}}{2}.
\end{align}
\end{subequations}

Eq.\eqref{eq:Simpson_like} is the Simpson-like quadrature in eq.\eqref{eq:Simpson_quadrature}, {  but there are a critical points in this derivation. If $x_\xi$ is a stochastic function, then it's nowhere differentiable. As a consequence, eqs.\eqref{eq:inquired_equalities} should not be valid. Similarly, the smoothing-approximation given by eqs.\eqref{eq:inquired_approximations} does not fit well when $x_\xi$ and $F_\xi$ are stochastic. In addition, eqs.\eqref{eq:inquired_approximations} break down the validity of the Simpson quadrature and introduce an uncontrolled error in eq.\eqref{eq:Simpson_Stieltijes}. As a consequence it's impossible to \emph{a priori} compare the results obtained with eq.\eqref{eq:Simpson_like} and the ones obtained with eq.\eqref{eq:Simpson_Stieltijes}, even in the case of deterministic functions.} All these observations tells us that we need to address both eqs.\eqref{eq:inquired_equalities} and eqs.\eqref{eq:inquired_approximations} for stochastic functions if we want to use this qualitative derivation of eq.\eqref{eq:Simpson_quadrature}.

\section{Possible extension to stochastic functions}\label{sec:stochastic_gen}
\begin{figure}
\includegraphics[width=\linewidth]{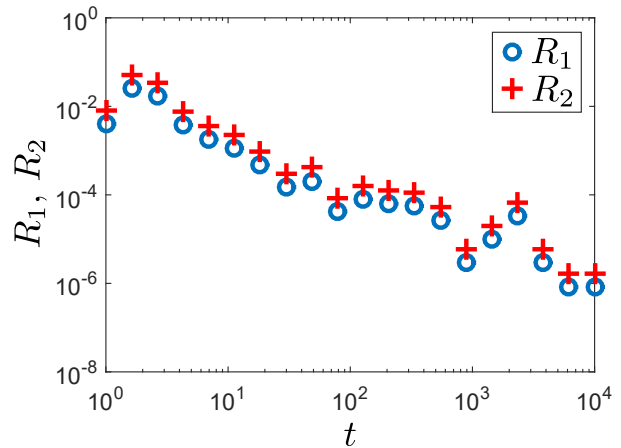}
\caption{Plot of $R_1$ and $R_2$ as functions of $t$.}\label{fig:integrals_inquire}
\end{figure}

\begin{figure}
\includegraphics[width=\linewidth]{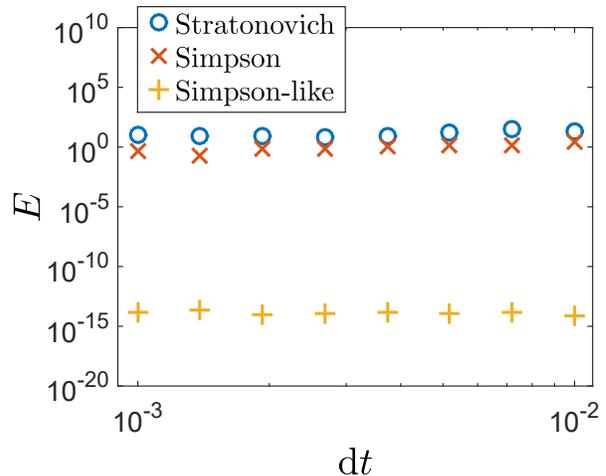}
\caption{Plot of $E$ for the different quadrature schemes in eq.\eqref{eq:Stratonivic_quadrature}, eq.\eqref{eq:Simpson_Stieltijes} and eq.\eqref{eq:Simpson_like}. }\label{fig:quality_inquire}
\end{figure}

To address the validity of eqs.\eqref{eq:inquired_equalities} and eqs.\eqref{eq:inquired_approximations} for stochastic functions we use different numerical simulations of a simple stochastic system. We take an overdamped colloidal particle that moves inside a quartic potential $H(x)=x^4$ {  under the action of a Gaussian white noise. Without the loss of generality, this system is described by the dimensionless stochastic differential equation
\begin{equation}\label{eq:Langevin}
\mathrm{d}x=-4\,x^3\, \mathrm{d}t+\sqrt{2d}\,\mathrm{d}W,
\end{equation}
where $\mathrm{d}W$ is a wiener process\footnote{ Equation \eqref{eq:Langevin} can be generalized so to include fluctuations with limited spectral-bandwidths. One generalization is
\begin{displaymath}
\begin{aligned}
\mathrm{d}x=&\left(-4\,x^3+u\right)\, \mathrm{d}t\\
\mathrm{d}u=&-\omega_c\, u\, \mathrm{d}t+\sqrt{2d}\,\omega_c\, \mathrm{d} W
\end{aligned}
\end{displaymath}
where $u$ is an Ornstein-Ulenbeck process with amplitude $d$ and cutoff pulsation $\omega_c$. The results obtained in this article are valid also for this latter case when $d=5$ and $\omega_c=50$. However, the size of  $\mathrm{d} t$  is  then important to interpret $u$ as a thermal bath \cite{mestres}. As a consequence, band-limited fluctuations requires more stringent conditions for $\mathrm{d} t$ than the ones for eq.\eqref{eq:Langevin}.} and $d$ is the dimensionless diffusion coefficient of the particle \cite{gardiner}. There are two reasons for this choice. The first one is that eq.\eqref{eq:Simpson_like} and eq.\eqref{eq:Stratonivic_quadrature} coincide when $F_\xi\propto x_\xi$. As a consequence, it's possible to see the advantages of eq.\eqref{eq:Simpson_like} over eq.\eqref{eq:Stratonivic_quadrature} only with non-quadratic potentials. The second reason is that many experiments with colloidal particles involve quartic potentials, especially the ones that address the Landauer principle and fluctuation theorems \cite{ciliberto,ciliberto2,roldan,gieseler,jun}. This means that the conclusions drawn with this simple system easily extend to all the experiments by which we address the foundations of stochastic thermodynamics today.  For the purposes of this article we will use $d=5$ and study the relaxation-to-equilibrium. To obtain this physical transformation we take $x_0$ from the nonequilibrium distribution $\mathcal{N}(0,0.01)$ and then we let the system freely evolve toward equilibrium with the Order 2.0 weak Taylor scheme  (O2WTS) \cite{klodenplaten}.

The first thing that we prove is the validity of eqs.\eqref{eq:inquired_equalities} for this system. To this end we take the largest timestep possible with the O2WTS  ($\mathrm{d}t=0.01$) and} we simulate the time evolution of $x_\xi$ between $\xi=0$ and $\xi=t$ for different values of $t$. The final value $x_t$ of this simulation is then used as the initial condition for a second simulation in which we evaluate $x_\xi$ between $\xi=t$ and $\xi=t+\mathrm{d}t$. The main difference with the first simulation is that the time-step of this second simulation is $\mathrm{d}\overline{t}=10^{-4}\,\mathrm{d}t$. This value is such that the numerical instabilities of the Stratonovich quadrature are not critical. As a consequence we can compute $I_1$ and $I_2$ with the Stratonovich quadrature as
\begin{subequations}\label{eq:inquired_equations_strat}
\begin{align}
I_1'\approx&\sum_{n=0}^{\frac{\mathrm{d}t}{\mathrm{d}\overline{t}}-1} \left[t+\left(n+\tfrac{1}{2}\right)\,\mathrm{d}\overline{t} \right]\left(x_{n+1}-x_n\right)\\
I_2'\approx&\sum_{n=0}^{\frac{\mathrm{d}t}{\mathrm{d}\overline{t}}-1} \left[t+\left(n+\tfrac{1}{2}\right)\,\mathrm{d}\overline{t} \right]^2\left(x_{n+1}-x_n\right).
\end{align}
\end{subequations}
where $x_n=x_{t+n\mathrm{d}\overline{t}}$ and the prime notation means that we are computing $I_1$ and $I_2$ with a different strategy than the one given in eq.\eqref{eq:inquired_equalities}. With eq.\eqref{eq:inquired_equations_strat} we can finally evaluate the relative errors
\begin{subequations}
\begin{align}
R_1=&\left|\frac{I_1'-I_1}{I_1'}\right|\\
R_2=&\left|\frac{I_2'-I_2}{I_2'}\right|,
\end{align}
\end{subequations}
between the right hand side of eq.\eqref{eq:inquired_equations_strat} and the one of eq.\eqref{eq:inquired_equalities}. Figure \ref{fig:integrals_inquire} shows that both $R_1$ and $R_2$ are decreasing functions of $t$: their maximum is approximately 0.01 and $R_{1,2}\to 0$ when $t\to \infty$. This gives us a numerical proof of eqs.\eqref{eq:inquired_equalities} when both $x_\xi$ and $F_\xi$ are stochastic processes.
{ As a consequence we can use eqs.\eqref{eq:inquired_equalities} to derive eq.\eqref{eq:Simpson_like} for stochastic functions, at least for the overdamped colloidal particle we are now considering here. However, this is not the main implication of Figure \ref{fig:integrals_inquire}. To our knowledge, Figure \ref{fig:integrals_inquire} is the first evidence in the literature where a deterministic integral-equality holds also for stochastic functions. This tells us that some quadrature algorithms may be valid for both deterministic and stochastic functions. Since the numerical integration of stochastic integrals has received little attention so far \cite{badr}, this result could give momentum to this topic in stochastic methods and stimulate the development of a unified framework to \emph{a priori} discuss the precision of approximations like eqs.\eqref{eq:inquired_equalities}-\eqref{eq:Simpson_like}. One possibility that we believe to be a promising starting point to approach this issue is to study eqs.\eqref{eq:I12defs} as stochastic integrals with respect to martingales \cite{kuo} while expanding $\mathrm{d}x_\xi$ as an Ito-Taylor/Stratonovich-Taylor differential \cite{klodenplaten}. This goes beyond the scope of this article and will not be discussed here.

Going back to the main aim of this paper we now show that the approximations eqs.\eqref{eq:inquired_approximations} are relevant for stochastic functions, even though they introduce an uncontrolled error}. The best strategy is to show that the Simpson-like formula eq.\eqref{eq:Simpson_like} performs similarly to the original Simpson formula eq.\eqref{eq:Simpson_Stieltijes}. This study is carried over as follows: we simulate the time evolution of $x_\xi$ from $\xi=0$ to $\xi=10$ for different sizes of the timesteps $\mathrm{d}t/2$. For each simulation we compute the cumulative heat exchanged with the reservoir between time $\xi=0$ and $\xi=t$ as
\begin{equation}\label{eq:cumulative_heat}
\Delta_t Q=\sum_{n=0}^{t/\mathrm{dt}} \int_{n\,\mathrm{d}t}^{n\,\mathrm{d}t+\mathrm{d}t} F_\xi \, \mathrm{d}x_\xi
\end{equation}
where the argument of the sum is evaluated with eq.\eqref{eq:Stratonivic_quadrature}, eq.\eqref{eq:Simpson_Stieltijes} \emph{and} eq.\eqref{eq:Simpson_like}. Since we have that for the relaxation to equilibrium process \cite{seifert,sekimoto,jarzynski2}
\begin{equation}
H(x_t)-H(x_0)+\Delta_t Q=0
\end{equation}
we can quantify the quality of the different quadrature schemes used in eq.\eqref{eq:cumulative_heat} through the absolute error
\begin{equation}
E=\max_{t<10}\left| H(x_t)-H(x_0) + \Delta_{t} Q \right|.
\end{equation}
Figure \ref{fig:quality_inquire} shows $E$  as a function of $\mathrm{d}t$. From the figure we see that the eq.\eqref{eq:Simpson_Stieltijes}-\eqref{eq:Simpson_like} yield $E$ values that respectively are 2 and 12 orders of magnitude smaller than the ones obtained with eq.\eqref{eq:Stratonivic_quadrature}. Overall, this is not surprising because three-points quadratures are expected to perform better than single-point ones \cite{numericalrecipes}. However there is an unexpected twist: the Simpson-like quadrature eq.\eqref{eq:Simpson_like} is much more precise than the original eq.\eqref{eq:Simpson_Stieltijes}. { Such an impressive precision-increase suggests that the error associated with eqs.\eqref{eq:inquired_approximations} may be more controlled than initially thought. We are not able to give a precise statement of this fact because we do not have a well defined formalism to address the \emph{a priori} precision of quadratures for stochastic integrals; nonetheless, we can surely conclude from Figure \ref{fig:quality_inquire} that the approximations in eq.\eqref{eq:inquired_approximations} seems to be necessary if we want to increase the precision of the three-points quadrature eq.\eqref{eq:Simpson_Stieltijes}.

The impressive performances of eq.\eqref{eq:Simpson_like} have two consequences} that are particularly important for experimental setups. The first one is that we do not have to trade time-sampling for precision in heat measurement: thanks to eq.\eqref{eq:Simpson_quadrature}, the same time-discretization is valid for both $x_\xi$ and $\Delta_\xi Q$. The same does not hold with eq.\eqref{eq:Simpson_Stieltijes} which instead implies that the time-discretization of $\Delta_\xi Q$ is coarser than the one of $x_\xi$. This is quite convenient in experiments as it allows to maximize the throughput of a single experimental run. The second important consequence is that eq.\eqref{eq:Simpson_like} gives a high accuracy even with very large $\mathrm{d}t$. Many areas in nonequilibrium thermodynamics could benefit from this feature. One example is the study of the Landauer principle, a topic where heat measurements needs to be extremely precise. This is usually achieved with fine sampling techniques \cite{bechhoefer2,ciliberto,ciliberto2,jun,gieseler,roldan,mestres} and state-of-the-art setups. Unfortunately, the knowledge required to build this kind of advanced setups may  not be immediately accessible to other researchers. Therefore, tools like eq.\eqref{eq:Simpson_like} could help to design simpler but effective setups and, thus, increase the number of experiments currently available in this research area. 

\section{Conclusions}\label{sec:conclusions}
In this paper we have addressed the validity of eq.\eqref{eq:Simpson_like} as numerical quadrature to evaluate the heat exchanged by a stochastic system. To this end we first derived eq.\eqref{eq:Simpson_like} when both $F_\xi$ and $x_\xi$ are deterministic functions. In our derivation we used two equations, eqs.\eqref{eq:inquired_equalities} and eqs.\eqref{eq:inquired_approximations}, which are critical because they should not hold for stochastic functions. Nonetheless, we were able to show that both eqs.\eqref{eq:inquired_equalities} and eqs.\eqref{eq:inquired_approximations} are well justified{ , at least for a stochastic system of experimental relevance. The main implications of this result are (i) that some integral equalities may hold for both deterministic and stochastic, and (ii) that it's possible to measure heat in real setups without state of the art sampling techniques. Clearly this latter feature is of capital importance for the field of stochastic thermodynamics as it means that it's possible to design reliable and sensitive experiments without state of the art instrumentation. 

To conclude this paper, we briefly comment an interesting feature that arises from eqs.\eqref{eq:inquired_approximations}.} Equations. \eqref{eq:inquired_approximations} exchange the real $x_\xi$ and $F_\xi$ with a smooth approximation. Since eqs.\eqref{eq:Simpson_like} yield an impressive improvement over the original quadrature eq.\eqref{eq:Simpson_Stieltijes}, we have a rather counter-intuitive implication: we can improve the performances of a stochastic quadrature scheme with a minor smoothing of the random process. {  The overall extent and numerical validity of this strategy are not clear, thus further investigations are of capital importance: in any experimental setup it's always more convenient to smooth a random signal than sample it with higher acquisition rates. As a consequence this counter-intuitive observation, if true in general,  may significantly help the design of new and more efficient quadrature schemes for stochastic integrals}. 

\begin{acknowledgments}
We would like to thank John Bechhoefer, Sergio Ciliberto, Luca Gammaitoni, M. L\'opez-Su\'arez and Igor Neri for the useful discussions on this topic. This work has been performed with the help of the study grant \emph{Energy transformation processes at the micro scales} from the USA Office of Naval Research.
\end{acknowledgments}

\bibliographystyle{unsrt}
\bibliography{riferimenti}

\end{document}